\documentclass[11pt]{article}

\pdfoutput=1

\usepackage[a4paper]{geometry}
\usepackage{graphicx}
\usepackage{amsmath}
\usepackage{amssymb}
\usepackage{color}
\usepackage{url}

\usepackage[subrefformat=parens,labelformat=parens]{subfig}

\def \conc {\varphi}					

\def \Vconc {\psi}				

\def \SPconc {\phi}

\def \temp {\Theta}					

\def \ds {\displaystyle}

\def \R{\mathbb{R}}

\def \M{\mathcal{M}}     		
\def \B{\mathcal{B}}     		
\def \O{\mathcal{O}}     		
\def \K{\mathcal{K}}

\def \W{\Omega}          		
\def \w{\omega}          		
     		
\def \iff{\Leftrightarrow}    		
    		
\def \to{\rightarrow}    		
\def \e{\varepsilon}             	
\def \he{\hat{\varepsilon}}             	
\def \E{\mathcal{E}}			
\def \I{\mathcal{I}} 			
\def \d{\delta}               		
\def \D{\Delta}               		
\def \p{\partial}               		
\def \ELL{\ell}
\def \usd{\mathrm{\text{ }d}}
\def \ud{\mathrm{d}}

\def \({\left(}				
\def \){\right)}

\begin{document}

\title{Step evolution in two-dimensional diblock copolymer films}

\author{Q. Parsons\thanks{Department of Computer Science, 
University of Oxford,
Wolfson Building,
Parks Road,
Oxford OX1 3QD, UK}
\and 
D. Kay\footnotemark[1]
\and 
A. M\"{u}nch\thanks{Mathematical Institute,
University of Oxford,
Andrew Wiles Building,
Woodstock Road,
Oxford OX2 6GG, UK}
}
\maketitle


\begin{abstract}
The formation and dynamics of free-surface structures, such as steps or
terraces and their interplay with the phase separation in the bulk are key
features of diblock copolymer films. We present a phase-field model with an
obstacle potential which  follows naturally from derivations of the
Ohta-Kawasaki energy functional via self-consistent field theory.  The free
surface of the film is incorporated into the phase-field model by including a
third phase for the void.  The resulting model and its sharp interface limit
are shown to capture the energetics of films with steps in two dimensions.  For
this model, we then develop a numerical approach that is capable of resolving the
long-time complex free-surface structures that arise in diblock copolymer films.
\end{abstract}

\section{Introduction}  \label{sec:intro}

Block copolymers are important designer materials with a range of important industrial and scientific applications. The monomers making up the molecules of these materials are organised into subchains or linear blocks of like monomers. In the simplest case, the diblock copolymer (DBC), each molecule consists of a linear block of type $A$ monomers connected by a covalent bond to a linear block of type $B$ monomers. More generally, repeated blocks, combinations with more than two monomers, and topologies other than linear chains are also possible. Block copolymer molecules usually prefer arrangements in which blocks of the same monomer species are close to each other. As such, they self-assemble into highly regular patterns of like monomers including, lamellae, cylinders or so-called gyroids \cite{materials_view}, by a process called phase separation. The size of the resulting segregated regions is limited by the fact that the blocks are chemically bonded, and by the tendency of the chains to coil up. Phase separation thus occurs on a micro- but not a macro-scale, making them extremely useful as functional nanomaterials \cite{kim2010functional}.

These polymer molecules can be precisely synthesized on a large scale to create diverse materials with properties that are determined by the structure of the molecules. Applications are, for example, in gel electrophoresis, where the lamellae are chemically tuned to have a preference for certain species in the material to be analysed \cite{lyakhova2006dynamics}. In other cases, the bulk properties that arise from the microstructures are exploited in commodity products such as box tape adhesives or as toughening additives for tarmacs \cite{materials_view}. For applications, it is important to understand, predict and ultimately control the microphase separation structures that arise in melts of specific block copolymer molecules. The self-consistent field theory (SCFT), which considers the average effect of the surrounding monomers for each molecule (see \cite{Edwar65, Genne69,Fredr05}), is an important theoretical tool for predicting structures and patterns in block copolymers. However, the mathematical analysis of these models is limited. Moreover, numerical methods based on the SCFT are computationally expensive and require the assumption of symmetries (see \cite{choksi_phases,UneyaD04}, and \cite{fredrickson2002field} for a review). In contrast, phase field or density functional theory (DFT) models \cite{OKFEF_original,Leibl80} of the Ohta--Kawasaki (or non-local Cahn--Hilliard) type are amenable to the analytical methods developed for Cahn--Hilliard type models. They also admit numerical approaches that do not require any solution symmetry assumptions. While it has been shown that they can be derived as approximations of the SCFT \cite{choksi_OKFEF_deriv,choksi2005diblock}, the validity of Ohta--Kawasaki models has been challenged, in particular in the intermediate and strong segregation regime, which is intensively studied in experiments. However, computations based on the DFT have reproduced all of the phases predicted by the SCFT, including gyroids \cite{choksi_phases}.

The effect of confining a polymer to a thin film, which arises frequently in applications, has attracted further attention, since the confinement introduces an additional length scale that is not typically compatible with the natural length scale selected by a layered pattern in free space. If, for example, a layer of a DBC is confined between two substrates with a gap that is highly incommensurate with the natural lamella period in the bulk, the system may switch to a state where the lamellae are perpendicular to the substrate, or a mixed state with both parallel and perpendicular interfaces \cite{Matse97, carvalho1994morphology}. If one of the surfaces is free, the system can also change thickness locally, thus creating steps and, in three dimensions, terraces. In three dimensions, microstructures other than lamellae can appear, such as cylinders of various orientations, or spheres, adding further complexity to the possible interactions between the structure in the polymer bulk and the shape of the free surface \cite{lyakhova2006dynamics,DijkB95,YokoyMK00}.

Carrying out computations based on the SCFT is costly; tracking terraces on large three-dimensional domains for long times (via dynamical variants of the SCFT) to observe the behaviour of surface structures, even more so. As such, the results in the literature are limited to a small number of carefully chosen domains \cite{lyakhova2006dynamics}. This limitation has renewed the interest in computations using DFT models \cite{pinna2012large}.  Our focus here is to explore the suitability of such an approach to capture step formation in two dimensions, and compare our results to those obtained using the SCFT in \cite{stasiak2012step}. We use an Ohta--Kawasaki type model \cite{OKFEF_original,UneyaD04} with an obstacle (rather than a polynomial) potential for the homogeneous free energy. This type of potential arises naturally in the SCFT formulation and in fact carries over to the DFT when the latter is derived systematically as an approximation to the SCFT \cite{choksi_OKFEF_deriv,choksi2005diblock}. The resulting model is attractive mathematically because, similar to the Cahn--Hilliard equation with double obstacle potentials \cite{BloweE91}, the exact the 1D lamellar equilibrium solutions can be stated explicitly (although the long-range interaction makes the derivations more challenging; see section~\ref{subsec:finite}). Moreover, away from the diffuse interface, the obstacle potential forces the phases to be pure. Efficient numerical methods are also available in the literature \cite{BloweE92,bosch2014fast}. We include a third, void or filler species to allow for a free surface of the DBC layer, following the example of similar SCFT simulations. A similar approach for films and droplets with a DBC and a solvent species was pursued in \cite{Cohen2014}, using the Brazovskii form for the free energy. This form uses a different representation of the long-range interaction arising from the covalent bond between copolymer blocks, and also a different homogeneous free energy.

The paper is structured as follows: in section \ref{sec:3phase_model}, we formulate the phase-field equation for a symmetric DBC with a void phase to model the free surface. In section \ref{sec:3phase_in_1D}, we explore the layered states with minimal energy when the film is flat, and use this to approximate steady states where the film has a step. In section~\ref{sec:num}, we present numerical solutions of the time-dependent Ohta--Kawasaki model to explore the dynamics of films with steps at the polymer--void interface. Section \ref{sec:conclusions} gives a summary of our results.

\section{Formulation}  \label{sec:3phase_model}
We consider a domain  $\W \subset \R^d$ ($d = 1,2,3$) with boundary $\p \W$ filled with a DBC comprising blocks of monomer species $A$  and $B$, and a third species $V$ for the void, which we model as a homopolymer. We restrict our attention to symmetric DBCs, where the $A$ and $B$ blocks are of equal length. The local proportions of the three species is represented by $\SPconc_A = \SPconc_A \( x \)$ (species $A$), $\SPconc_B = \SPconc_B \( x \)$ (species $B$) and $\SPconc_V = \SPconc_V \( x \)$ (the void), $x\in\Omega$.  The void $V$ acts as a filler to allow the copolymer to have a free surface. Each of these variables is bounded by $0$ and $1$. Together, they satisfy the incompressibility condition
\begin{align}  \label{eq:3phase_incompressibility}
\SPconc_A + \SPconc_B + \SPconc_V = 1,
\end{align}
since the two copolymer species and the void fill the space. Hence, $\SPconc_i \( x \) = 0$ implies no phase $i$ material whatsover at $x$, and $\SPconc_i \( x \) = 1$ implies pure phase $i$ material (and nothing else) at $x$. We denote the average of any scalar function $f$ over $\W$ as
\[
\overline{f} := \dfrac{1}{|\W|} \int_{\W}{f \usd \W},
\]
where $|\W|$ denotes the Lebesque measure of $\Omega$. Incompressibility \eqref{eq:3phase_incompressibility} and symmetry of the copolymer molecules imply that
\begin{align}  \label{eq:avrel}
\overline{\SPconc_A} + \overline{\SPconc_B} + \overline{\SPconc_V} &= 1,
& &\text{where} &
\overline{\SPconc_A} &= \overline{\SPconc_B},
\end{align}
since each molecule consists of equal numbers of $A$ and $B$ monomers, which in turn means that the total numbers of $A$ and $B$ monomers are equal.

Our goal is to investigate the behaviour of a DBC which has a free boundary with the surrounding void, subject to surface tension. We do so by minimising a free energy of the form (cf.~(3.22) of \cite{choksi2005diblock}) 
\begin{align}   \label{eq:generic_3_phase_free_energy}
\E \( \boldsymbol{\SPconc} \)  = \int_{\W} \I \( \nabla \boldsymbol{\SPconc} \) \usd \W + \int_{\W}{ \mathcal{N} \( \boldsymbol{\SPconc} \) \usd \W} + \int_{\W}{ \B \( \boldsymbol{\SPconc} \) \usd \W},
\end{align}
comprising the sum of interfacial, non-local and homogeneous free energy contributions. Here, we have introduced $\boldsymbol{\SPconc} := \( \SPconc_A, \SPconc_B, \SPconc_V \) = \( \SPconc_1, \SPconc_2, \SPconc_3 \)$, and will switch between the two types of subscripting as convenient. According to \cite{choksi2005diblock,choksi_OKFEF_deriv}, the first two terms of 
\eqref{eq:generic_3_phase_free_energy} comprise the entropic part of the free energy, while the last term is the enthalpy.

The interfacial energy is
\begin{align}  \label{eq:3_phase_interfacial_energy}
\I \( \nabla \boldsymbol{\SPconc} \) = \dfrac{1}{2} \sum_{i=1}^{3}{ \hat K_i \he^2 \left| \nabla \SPconc_i \right|^2 },
\end{align}
where $\he \ll 1$. Working from the matrix $K$ in (3.24) of \cite{choksi_OKFEF_deriv}, we set $\hat K_{1} = \hat K_{2}$ for a symmetric DBC.

The second, long-range term in the free energy \eqref{eq:generic_3_phase_free_energy} models the fact that the $A$ and $B$ blocks are chemically bonded and resist separating, even though the individual $A$ and $B$ monomers repel each other. We use
\begin{align}  \label{eq:3_phase_long_range_energy_simple}
\mathcal{N} \( \boldsymbol{\SPconc} \) &= \dfrac{1}{2} \sum_{i,j=1}^{2}{\he\hat\gamma_{ij} \( \( - \D_n \)^{-1} \( \SPconc_i - \overline{\SPconc_i} \) \) \( \SPconc_j - \overline{\SPconc_j} \)},
\end{align}
where the shorthand `$\( - \D_n \)^{-1} f$' denotes the solution, $\Phi$, of Poisson's equation with zero Neumann boundary conditions,
\begin{align}  \label{eq:3_phase_long_range_energy_simple_aux_eqn}
- \D \, \Phi &= f, & \left. \p_n \Phi \right|_{\p \W} &= 0,  & \overline{f} &= \overline{\Phi} = 0.
\end{align}
The `$_n$'-subscript signifies the zero Neumann boundary condition. A Green's function formulation allows us to solve for each $\( - \D_n \)^{-1} \( \SPconc_i - \overline{\SPconc_i} \)$ as
\begin{align*}
\( - \D_n \)^{-1} \( \SPconc_i - \overline{\SPconc_i} \) \( x \)&= \int_{\W}{ G \( x, y \) \( \SPconc_i \( y \) - \overline{\SPconc_i} \) \ud y}
\end{align*}
as in \cite{BloweE91}. Note that \eqref{eq:3_phase_long_range_energy_simple_aux_eqn} and our operator $\( - \D_n \)^{-1}$ are well-defined, since $\overline{f} = \overline{\Phi} = 0$.

For the $\hat \gamma_{ij}$ coefficients in \eqref{eq:3_phase_long_range_energy_simple}, we use the matrix $J$ in (3.25) of \cite{choksi2005diblock}, and our assumption of molecular symmetry 
to write
\begin{equation}  \label{eq:3phase_nonlocal_matrix_choksi}
\hat\gamma_{ij}=(-1)^{i+j} \hat\gamma,
\end{equation}
with $\hat\gamma$ a constant parameter. In restricting the sum \eqref{eq:3_phase_long_range_energy_simple} to $i=1,2$, we have explicitly assumed that there is no long-range interaction between the two copolymer species and the void.

The third term, $\B$, is the homogeneous contribution to the free energy, and accounts for the short-range interaction between the three species. We use an obstacle potential that naturally appears in the derivation of Ohta--Kawasaki type DFT models from SCFT models. As in \cite{choksi2005diblock,choksi_OKFEF_deriv}, we use
\begin{align}  \label{eq:3phase_choksi_density}
\B \( \boldsymbol{\SPconc} \) = \begin{cases}
\ds \dfrac{1}{2 \temp} \( \sum_{k,m \in \left\{ A, B, V \right\}}{\hat\B^{km} \SPconc_k \SPconc_m} - \sum_{k \in \left\{ A, B, V \right\}}{\hat\B^{kk} \SPconc_k} \) & \sum{\SPconc_k} = 1; \SPconc_k \geq 0 \\
\infty & \text{otherwise,}
\end{cases}
\end{align}
where $\temp$ denotes the absolute temperature. We immediately conclude that $\B \( \boldsymbol{\SPconc} \) \geq 0$ for all $\boldsymbol{\SPconc}$ and $\B \(\boldsymbol{\SPconc} \) = 0$ $\iff$ $\boldsymbol{\SPconc} = \( 1, 0, 0 \)$ or $\( 0, 1, 0 \)$ or $\( 0, 0, 1 \)$ i.e.~for pure phases. The coefficients ${\cal B}^{km}$ are related to the three Flory--Huggins parameters \cite{choksi2005diblock} via
\begin{subequations}  \label{eq:flory_huggins_from_V_interaction}
\begin{align}
\hat\chi^{AB} &= \dfrac{1}{\temp} \( \hat\B^{AB} - \dfrac{1}{2} \( \hat\B^{AA} + \hat\B^{BB} \) \) > 0,  \label{eq:flory_huggins_from_V_interaction_AB} \\
\hat\chi^{AV} &= \dfrac{1}{\temp} \( \hat\B^{AV} - \dfrac{1}{2} \( \hat\B^{AA} + \hat\B^{VV} \) \) > 0,  \label{eq:flory_huggins_from_V_interaction_AC} \\
\hat\chi^{BV} &= \dfrac{1}{\temp} \( \hat\B^{BV} - \dfrac{1}{2} \( \hat\B^{BB} + \hat\B^{VV} \) \) > 0.  \label{eq:flory_huggins_from_V_interaction_BC}
\end{align}
\end{subequations}
Notice that in each case, the repulsion between each pair of unlike material phases exceeds the average repulsion of the corresponding pairs of like phases.

As in \cite{ohta1995dynamics}, \cite{ito1998domain} and Appendix A of \cite{choksi2005diblock}, we use the incompressibility condition to reduce the number of unknowns by one. We define a \emph{micro}-phase separation parameter $\conc$, and a \emph{macro}-phase parameter $\Vconc$, in which
\begin{align}  \label{eq:two_order_parms}
\conc &= \SPconc_A - \SPconc_B, & \Vconc &= \SPconc_A + \SPconc_B,
\end{align}
in which $\conc = 1$ ($-1$) denotes pure phase $A$ ($B$). Combining these with incompressibility in \eqref{eq:generic_3_phase_free_energy}, leads to the free energy
\begin{subequations}\label{eq:ropm}
\begin{align}\label{eq:ropma} 
\E \( \conc, \Vconc \)  = \int_{\W}{ \dfrac{\e^2}{2} \left( \K \left| \nabla \Vconc \right|^2 +  \left| \nabla \conc \right|^2 \right)
+ \dfrac{\e\gamma}{2} \( \conc - \overline{\conc} \) \left( \( - \D_n \)^{-1} (\conc-\overline{\conc})\right) + \B \( \conc, \Vconc \)  \ud \W},
\end{align}
for a symmetric DBC, in which
\begin{equation} 
\B \( \conc, \Vconc \) = \dfrac{1}{2} \left[ \( \dfrac{1}{2} - 2\chi^{AV} - \chi^\Delta \) \Vconc^2 - \dfrac{1}{2} \conc^2 + \chi^\Delta \conc (\Vconc-1)
+ \(2 \chi^{AV} + \chi^\Delta\) \Vconc  \right], 
\end{equation}
\end{subequations}
for $|\conc|\leq \Vconc$ and $0 \leq \Vconc \leq 1$, and $\infty$ otherwise. Notice that we have rescaled $\E$ with $1/\hat\chi^{AB}$, and so the parameters are
$\e^2 = \hat K_1 {\he^2}/{2\hat \chi^{AB}} $, 
$\K= 1 + 2 \widehat{K}_3/\widehat{K}_1 = 1 + 2 \widehat{K}_3/\widehat{K}_2$,
and $\e\gamma=\he\hat\gamma/\hat \chi^{AB}$,
$\chi^{AV}=\hat\chi^{AV}/\hat \chi^{AB}$,
$\chi^{\Delta}=(\hat\chi^{BV}-\hat\chi^{AV})/\hat \chi^{AB}$.

\subsection*{Terminology}
We adopt the following terminology in what follows, and refer to Fig.~\ref{fig:terracing_schematic} below and Fig.~1 of \cite{stasiak2012step} for details. A $BA$ molecular layer or monolayer corresponds to a single molecule, comprising a section of (nearly) pure phase $B$ material ($\conc \approx -1$), a $BA$ internal interface layer from $\conc = -1$ to $\conc = 1$, and a section of (nearly) pure phase $A$ material ($\conc \approx 1$), and similarly for an $AB$ monolayer. We study configurations in which the molecules are aligned perpendicular to a substrate, forming layers of $A$ and $B$-rich polymer material that are parallel to the substrate. We represent these layered solutions with sequences such as $ABBAA \ldots BBAV$, perpendicular to the substrate. The transition from the polymer to the void is either an $AV$ or $BV$ polymer--void interface layer. Given a function $\conc$ representing such a configuration, the $A$ and $B$-microdomains refer to those parts of the domain where $\conc = \pm 1$. Notice in the sequence above that the first (at the substrate) and last (just before the void) pure polymer phase segments (and the corresponding microdomains), are half as wide as the others. A thin film of $N$ monolayers stacked parallel to a substrate thus comprises $N+1$ $A$ or $B$-microdomains. 
\subsection*{Gradient flow dynamics}
We wish to explore the dynamics numerically. For this purpose we introduce the Moreau--Yosida regularisation of the free energy 
\begin{subequations}
\begin{align}\label{eq:ropmmu} 
\E_\mu  = \int_{\W}{ \dfrac{\e^2}{2} \left(\K \left| \nabla \Vconc
\right|^2 +  \left| \nabla \conc \right|^2 \right) + \dfrac{\e\gamma}{2} \conc \left( \( - \D_n \)^{-1}  (\conc-\overline{\conc})\right) + \B_\mu \( \conc, \Vconc \)  \usd \W},
\end{align}
where 
\begin{align}\label{eqn:defbmu} 
\B_{\mu} \( \conc, \Vconc \) &:= \B \( \conc, \Vconc \) + \dfrac{1}{2\mu} \left[ ( \Vconc - 1 )_+^2 + (\conc - \Vconc )_+^2 + ( \conc + \Vconc )_-^2 \right],
\end{align}
\end{subequations}
for $0 < \mu \ll 1$. This generalises a choice made in \cite{bosch2014fast, hintermuller2011adaptive} for the Cahn--Hilliard scalar phase field model. The notation $(\cdot)_+$ and $(\cdot)_-$ indicates the positive and negative part of the expressions in the brackets. The homogeneous free energy $\B_{\mu}$ is defined for all values of $\conc$ and $\Vconc$ and tends to $\B$ as $\mu \downarrow 0$, that is, to the obstacle free energy.
 
We obtain the evolution of the system from the energy functional \eqref{eq:ropmmu} by initially using the $H^{-1}$-gradient flow (see Appendix B of \cite{choksi_phases}) to specify that
\begin{subequations} \label{eqn:timedep} 
\begin{align}  \label{eq:timedep_1st_attempt}
\partial_t\Vconc &= \nabla \cdot \( M_{\Vconc} \nabla \dfrac{\d \E_\mu}{\d \Vconc} \),
& &\text{and} &
\partial_t\conc &= \nabla \cdot \( M_{\conc} \nabla \dfrac{\d \E_\mu}{\d \conc} \),
\end{align}
for $t \in \( 0, T \right]$. We choose mobilities $M_{\conc} = 2$ and $M_{\Vconc} = \frac{2}{3}$, which is equivalent to using unit mobilities in the gradient flow formulation for $\SPconc_A$, $\SPconc_B$, and $\SPconc_V$. The first variation of $\E_{\mu}$ is given by
\begin{align}
\dfrac{\d \E_\mu}{\d \Vconc} &= \K \e^2 \( - \D \) \Vconc + \p_{\Vconc} \B_{\mu} \( \conc, \Vconc \), \label{eq:3phase_1st_variation_macro} \\
\dfrac{\d \E_\mu}{\d \conc} &= \e^2 \( - \D \) \conc + \p_{\conc} \B_{\mu} \( \conc, \Vconc \) + \e\gamma \( - \D_n \)^{-1} \( \conc - \overline{\conc} \). \label{eq:3phase_1st_variation_micro}
\end{align}
To reduce notational burden, we do not supply $\Vconc$ and $\conc$ above with an index $\mu$ to indicate that they solve the regularised problem, and will only do so where necessary. We use the natural (zero Neumann) boundary conditions on $\conc$ and $\Vconc$ from the first variation, and ensure global mass conservation by additionally imposing zero flux at the boundary for each of, 
\begin{align} 
\left. \p_n \conc \right|_{\p \W} &= 0, & \left. \p_n \D \conc \right|_{\p \W} &= 0,
\label{eq:regtime:bc1}\\
\left. \p_n \Vconc \right|_{\p \W} &= 0, & \left. \p_n \D \Vconc \right|_{\p \W} &= 0.
\label{eq:regtime:bc2}
\end{align}
Finally, we impose the initial conditions
\begin{align}\label{eqn:phipsi-ic}  
\Vconc \( x, 0 \) &= \Vconc_0 \( x \), & \overline{\Vconc_0} &= \mathcal{M}, & &\text{and} & \conc \( x, 0 \) &= \conc_0 \( x \), & \overline{\conc_0} &= 0,
\end{align}
\end{subequations}
where $\mathcal{M}$ denotes the prescribed copolymer volume fraction, $0 < \mathcal{M} < 1$ and we set $\overline{\conc_0} = 0$ to reflect the symmetry of the DBC molecules. Under these dynamics, the energy, $\E_{\mu} = \E_{\mu} \( t \)$, is a non-increasing function of time, which we demonstrate informally as follows: we use our scalar mobilities to re-arrange \eqref{eq:timedep_1st_attempt} as 
\begin{align*}
\dfrac{\d \E_\mu}{\d \Vconc} &= \dfrac{-1}{M_{\Vconc}} \( - \D_n \)^{-1} \partial_t \Vconc
& &\text{and} &
\dfrac{\d \E_\mu}{\d \conc} &= \dfrac{-1}{M_{\conc}} \( - \D_n \)^{-1} \partial_t \conc.
\end{align*}
Then
\begin{align*}
\dfrac{\ud \E_{\mu}}{\ud t} 
= \int_{\W}{ \dfrac{\d \E_{\mu}}{\d \conc} \p_t \conc + \dfrac{\d \E_{\mu}}{\d \Vconc} \p_t \Vconc \usd \W} = - \( \dfrac{\left\| \partial_t \Vconc \right\|^2_{-1}}{M_{\Vconc}} + \dfrac{\left\| \partial_t \conc \right\|^2_{-1}}{M_{\conc}} \) \leq 0,
\end{align*}
where the $\left\| \cdot \right\|_{-1}$ norm comes from (2.4) of \cite{BloweE91}. The system is also mass conserving, in the sense that $\overline{\Vconc} \( t \) = \overline{\Vconc_0} = \mathcal{M}$ and $\overline{\conc} \( t \) = \overline{\conc_0} = 0$ for every $t$. We therefore replace the non-local $\overline{\conc}$ in \eqref{eq:3phase_1st_variation_micro} with zero, and compute with the pair
\begin{subequations}  \label{eq:3phase_1st_variation_dynamics_as_used}
\begin{align}
\Vconc_t &= M_{\Vconc} \D \w_{\Vconc}
& &\text{where} &
\w_{\Vconc} &= \K \e^2 \( - \D \) \Vconc + \p_{\Vconc} \B \( \conc, \Vconc \); \\
\conc_t &= M_{\conc} \( \D \w_{\conc} - \e \gamma \conc \)
& &\text{where} &
\w_{\conc} &= \e^2 \( - \D \) \conc + \p_{\conc} \B \( \conc, \Vconc \).
\end{align}
\end{subequations}
These dynamics \emph{conserve} $\overline{\Vconc} \( t \)$, but \emph{control} $\overline{\conc} \( t \)$, in the sense that $\overline{\conc} \( t \) \to 0$ exponentially fast if $\overline{\conc_0} \neq 0$ and $\gamma > 0$. This property is useful for numerical simulations.

\section{Analysis of layered solutions}  \label{sec:3phase_in_1D}
\subsection{Sharp interface limit}\label{subsec:sharp}
We investigate 2D thin films of DBCs where the copolymer is organised in layers oriented parallel to the substrate (except for steps in the free surface), where the monolayer width is close to optimal. We first consider 1D configurations that are vertical sections through 2D solutions that are constant parallel to the substrate, in the limit as $\e \to 0$. As in Section 4 of \cite{choksi2005diblock}, we consider a film lying on a substrate at $y=0$, occupying $0 < y <\M L$, comprised of an alternating sequence of $A$ and $B$ microdomains. The rest of the domain, $\M L < y < L$, is void. We then consider the situation where the number of monolayers varies laterally in 2D, yielding a step-like change in the film thickness.

The optimal layered pattern on a fixed 1D domain $0<y<L$, minimises the energy $\E$ under the boundary conditions \eqref{eq:regtime:bc1} and \eqref{eq:regtime:bc2}, that is $\left. \conc' \right|_{0, L} = \left. \conc''' \right|_{0, L} = 0$ and $\left. \Vconc' \right|_{0, L} = \left. \Vconc''' \right|_{0, L} = 0$. To identify the scaling for the optimal number of copolymer monolayers $N$, and hence their optimal width $w_m = \M L/N$, we need to balance the energy contribution from the $AB$/$BA$ interface layers, which \emph{grows} with $N$, against that from the long-range interactions due to the chemical bonds, which \emph{decays} with $N$. More specifically, the energy from a single $AB$ or $BA$ interface layer scales like the gradient term in \eqref{eq:ropma}. Since the width of the diffusive layer $\O \( \e \)$ scales like $\e$, the gradient scales like $1/\e$, and the gradient term integrand in \eqref{eq:ropma} is $\O \( 1 \)$ inside the interface layer; outside of this, it does not contribute to the energy. Likewise, the homogeneous energy is $\O \( \e \)$ inside each interface layer. Hence the total (local) energy contribution from $N$ interface layers is $\sim N \e$. The contribution from the long-range (non-local) interaction comes from the second term in \eqref{eq:ropma}.  For a single monolayer of width $w= \M L/N$, the integrand scales like $\e\gamma w^2$ and contributes to the energy over a region of width $w$ (i.e.~the entire width of the monolayer, not just the interface layer). Hence the total long-range contribution is $\sim  N\e\gamma w^3$. Balancing these gives the optimal width $w_m \sim \gamma^{-1/3}$. To keep the width and number of monolayers on $[0,\M L]$ constant for an optimal configuration, we therefore consider the sharp interface limit in which we keep $\gamma$ fixed as $\e \to 0$.

In the sharp interface limit $\e \to 0$, $\gamma, \K$ fixed, the rescaled 1D free energy $\I_{\e} = \E/\e$ (with $\Vconc = 1$) tends to the functional \cite{Baldo1990,choksi2005diblock}
\begin{align}  \label{eq:I0}
\I_0={\sigma}_{AB}N+\sigma_{AV}N_{AV}+\sigma_{BV}N_{BV} +\frac{\gamma}2\int_{0}^{\M L}{\( \int_0^y \conc(z) \usd z \)^2 \ud y},
\end{align}
defined on the set
\begin{align*}
\left\{ \( \conc, \Vconc \): \left[ 0, L \right] \to \left\{ \( \pm 1, 1 \), \( 0, 0 \) \right\};\, \conc, \Vconc \in \mathcal{BV} \( \left[ 0, L \right] \),\ \overline\conc=0, \, \overline \Vconc= \M \right\},
\end{align*}
and the minimisers of $\I_{\e}$ tend to the minimisers of $\I_0$. Here, $\mathcal{BV} \( \left[ 0, L \right] \)$ denotes the set of functions with bounded variation, and $N$, $N_{AV}$, $N_{BV}$ the number of $AB$, $AV$, and $BV$ interface layers. The surface tension of each of these interface layers is given by
\begin{subequations}\label{eqn:surface-tension-values}
\begin{align}
\sigma_{AB} &= \inf_{\eta} \left\{ \varrho \left[ \eta \right]; \, \eta \in C^1, \, \eta \( 0 \) = \( -1,1 \),\, \eta \( 1 \) = \( 1,1 \) \right\} = \dfrac{\pi}{2 \sqrt{2}}, \\
\sigma_{AV} &= \inf_{\eta} \left\{ \varrho \left[ \eta \right]; \, \eta \in C^1, \, \eta \( 0 \) = \( 1,1 \),\, \eta \( 1 \)= \( 0,0 \) \right\} = \dfrac{\pi \sqrt{ \( 1 + \K \) \chi^{AV} }}{4 \sqrt{2}}, \\
\sigma_{BV} &= \inf_{\eta} \left\{ \varrho \left[ \eta \right]; \, \eta \in C^1, \, \eta \( 0 \) = \( -1,1 \),\, \eta \( 1 \) = \( 0,0 \) \right\} = \dfrac{\pi \sqrt{ \( 1 + \K \) \( \chi^{AV} + \chi^{\D} \)}}{4 \sqrt{2}},
\end{align}
where we have used the abbreviation (cf.~(4.35) of \cite{choksi2005diblock})
\begin{align}
\varrho \left[ \eta \right] := \sqrt{2} \int_{0}^{1}{ \sqrt{ \B \( \eta \( t \) \) \( \( \eta_1' \)^2 + \K \( \eta_2' \)^2 \) } \usd t }.
\end{align}
\end{subequations}
In \cite{choksi2005diblock}, the authors show that the microdomains of solutions with an alternating $\ldots AB \ldots $ pattern, which are critical points of $\I_0$, have the same width $w = \M L / N$, except for those adjacent to the substrate and the free surface, which are only half as wide. Consequently, the contribution from the last term in \eqref{eq:I0} to $\I_0$ is $N$ times the contribution from a single monolayer which extends symmetrically to both sides of an $AB$ or $BA$ interface layer. The first layer, in particular, starts at the substrate, $y=0$, and ends at $y = \M L/N$ and has an internal interface layer centered at $y = \M L/2N$. 
Hence, the contribution of the last integral term in \eqref{eq:I0} to $\I_0$ is
\begin{align*}
N \dfrac{\gamma}{2} \int_{0}^{w}{ \( \int_{0}^{x}{ \conc \( \xi \) \ud \xi } \)^2 \ud x } = N \dfrac{\gamma}{2} \int_{0}^{w}{ \( \left| x - \dfrac{w}{2} \right| - \dfrac{w}{2} \)^2 \ud x } =
N \dfrac{\gamma w^3}{24}.
\end{align*}
Evaluating $\I_0$ for such a 1D pattern thus gives
\begin{multline}  \label{eq:i0nm}
I_0 \( N, \M \) := \dfrac{\pi}{2 \sqrt{2}} N + \dfrac{\gamma}{24} \dfrac{\M^3 L^3}{N^2} \\ + \dfrac{\pi}{4\sqrt2}\sqrt{ 1 + \K } \left[ \sqrt{\chi^{AV}} N_{AV} +\sqrt{\( \chi^{AV} + \chi^{\D} \)} N_{BV} \right],
\end{multline}
where we have emphasised the dependence of $I_0$ on $N$ and the copolymer volume fraction, $\M$. We assume there is one polymer--void interface layer ($AV$ or $BV$), so either $N_{AV}=1$ and $N_{BV}=0$, or vice versa.  To determine the microdomain width giving the smallest $I_0$,  we replace $N$ by $\M L/w$ and minimise with respect to $w$, to obtain
\begin{align}  \label{eqn:wmin-si}
w_m = \dfrac{\( 3 \sqrt{2} \pi \)^{1/3}}{\gamma^{1/3}} = \dfrac{2.37}{\gamma^{1/3}},
\end{align}
which is independent of $\M$ and $L$ (here and elsewhere, we round irrational numbers to the number of digits given). On the other hand, $\M L/w_m$ is not generally an integer, and so the {actual} (integer) value for the layer count $N$, is whichever of $\left\{ \lfloor \M L/w_m \rfloor, \lceil \M L/w_m \rceil \right\}$ yields the smaller $I_0$ (both, if the value is the same).

\subsubsection*{2D stationary solutions with steps}
\begin{figure}[tp]
\begin{center}
  \subfloat[An $L_1^{||}$--$L_3^{||}$ configuration. The step is at $x = \xi L_x$.]{\label{fig:L_13_terrace}\includegraphics[clip=true,width=0.45\textwidth]{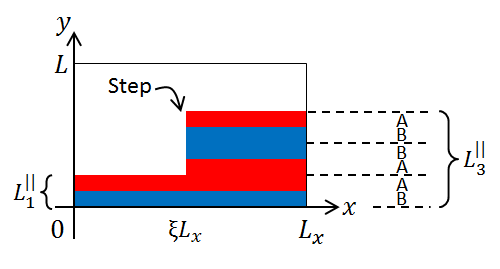}} \hspace{10pt}
  \subfloat[The corresponding flat $L_2^{||}$ configuration, $BAABV$.]{\label{fig:L_2_flat}\includegraphics[clip=true,width=0.45\textwidth]{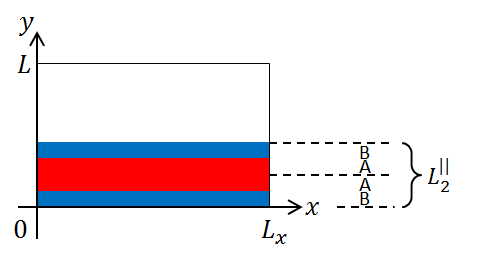}}
\caption{\label{fig:terracing_schematic}2D Step schematic. The $A$-microdomains are red; the $B$-microdomains blue. The microdomains closest to the substrate ($y=0$) and void are half as wide as the others. Both configurations have the same copolymer volume fraction, $\M$.}
\end{center}
\vspace{-10pt}
\end{figure}

We consider a 2D domain $[0,L_x]\times [0,L]$, where $0<x<L_x$ is the coordinate along the substrate, and $0<y<L$ is normal to it. We suppose that the domain is occupied by a (pure) copolymer of area $V_p = \M L \times L_x$ and a void with area $\( 1 - \M \) L \times L_x$. A schematic of the setup is depicted in Fig.~\ref{fig:terracing_schematic}.

A flat (uniform) copolymer film could thus occupy the area $0 < y < \M L$, and comprise $N$ layers parallel to the substrate, with $N$ as specified immediately after \eqref{eqn:wmin-si}. If, however, $w_m$ is not commensurate with an integer number of monolayers, the copolymer film could attain a non-uniform profile as an alternative to being flat (with monolayers of sub-optimal width $w \neq w_m$). The film surface could adjust to accomodate different numbers of layers on different parts of the substrate, all of optimal width $w_m$. The simplest possibility is a single step change in the film thickness at $x=\xi L_x$, $0<\xi<1$, at which the number of $AB$/$BA$-interface layers changes from $N_1$ (each of width $w_1$) to
$N_2$ (each of width $w_2$).  If we ignore the energy due to the step itself, the total energy is 
\begin{align}  \label{eq:generic_2D_stepped_energy_est}
I_{2D} = \xi L_x I_0 \( N_1, N_1 w_1 \) + \( 1 - \xi \) L_x I_0 \( N_2, N_2 w_2 \),
\end{align}
with $I_0$ as in \eqref{eq:i0nm}. We assume that the entirety of the top of the film surface (on both sides of the step) is made up of the same $A$ or $B$ species i.e.~one of $N_{AV}$ or $N_{BV}$ is non-zero for both contributions to $I_{2D}$. For a given copolymer fraction $\M$, we must also satisfy the mass constraint
\begin{align*}
\xi N_1 w_1 + (1-\xi) N_2 w_2= \M L.
\end{align*}
$I_{2D}$ is minimised when $w_1 = w_2 = w_m$, i.e.~the monolayers have the same width as for the 1D minimum. Under these conditions, the constraint implies that
\begin{align}  \label{eqn:steploc}
\xi = \dfrac{N_2-\M L/w_m}{N_2-N_1}.
\end{align}
We focus on the case where $N_1=N$, $N_2=N+2$, $\M L = \( N+1 \) w_m$.  Then $\xi=1/2$ and if the microdomain adjacent to the void is of type $A$, say, then the lateral energy density (in $x$) of the $L_{N}^{||}$--$L_{N + 2}^{||}$ configuration is 
\begin{align}  \label{eq:step_energy_sharp_interface_simple}
\dfrac{I_{2D}}{L_x} = \dfrac{3\pi}{4 \sqrt {2}} \( N+1 \) + \dfrac{\pi}{4 \sqrt{2}} \sqrt{\( 1 + \K \) \chi^{AV}}.
\end{align}
In this case, an alternate flat $L_{N+1}^{||}$ configuration with $N+1$ microdomains of optimal width $w_m$ is possible, with a $B$-type microdomain at the polymer--void interface layer, and lateral energy density
\begin{align}  \label{eq:flat_energy_sharp_interface_simple}
I_0 \( N+1, \M \) = \dfrac{3 \pi}{4 \sqrt{2}} \( N+1\) + \dfrac{\pi}{4 \sqrt{2}} \sqrt{ \( 1 + \K \) \( \chi^{AV} + \chi^{\D} \)}.
\end{align}
In this simple, sharp interface approximation, it is clear that if $\chi^{\Delta}>0$, then the configuration with the step has lower energy \eqref{eq:step_energy_sharp_interface_simple} than the flat, $\( N+1 \)$-layer alternative \eqref{eq:flat_energy_sharp_interface_simple}. Since this approximation neglects the energy due to the step itself, we expect that $\chi^{\Delta}$ should be above some threshold for the stepped configuration to be energetically advantageous (for a fixed $L_x$). On the other hand, the third term in $I_0$ in \eqref{eq:i0nm}, which measures the energy in the polymer--void interface layer, does not depend on $N$ (or $w$), and thus has no impact on $w_m$ in \eqref{eqn:wmin-si}. This simplification thus entirely discounts the effect of the polymer species at the void interface layer, on the layer count. Specifically, for very high $\chi^{\D}$, we might expect a nett energetic advantage from a higher/lower monolayer count (i.e.~monolayers of width other than $w_m$) if this puts the lower energy polymer species at the polymer--void interface layer. In the example above, we thus anticipate at least three possibilities: i) a flat film with $N+1$ layers of width $w_m$, but a very high energy polymer--void interface layer, ii) a flat film with $N$ or $N+2$ layers and a low energy polymer--void interface layer, but layer width not close to $w_m$, and iii) a configuration with $N$ and $N+2$ layers to the left/right of a step, with the lower energy polymer--void interface layer covering at least the majority of the film, and all of the molecular layers having width at or near $w_m$. Clearly, the relative advantage of this last configuration grows with increasing $L_x$. This suggests that stepped configurations should inevitably become energetically favoured for sufficiently large $L_x$, compared to the flat alternatives discussed above. We explore these issues in our numerical experiments in section \ref{sec:num}.

\subsection{Finite $\e$}  \label{subsec:finite}

In the numerical section \ref{sec:num},  we will choose our parameters to study a specific geometric scenario, and aim to ensure that the optimal monolayer spacing fits with these constraints. In view of the sharp interface results in section~\ref{subsec:sharp}, this amounts to choosing $\gamma$; there is usually more flexibility in the choice of $\e$. Small $\e$ will make the interface layers thinner and improve the accuracy of the sharp interface estimates, but increase the computational cost by requiring finer spatial grids. It is therefore useful to explore the effect of $\e$ on layered solutions in more detail. The use of the obstacle potential allows us to express the interface layers exactly with trigonometric functions. This is attractive from an analytic point of view, as was demonstrated in \cite{BloweE91,cahn1996cahn} for the example of the Cahn--Hilliard equation (equivalent to the current model with $\gamma=0$).

\begin{figure}[t]
\centerline{\includegraphics[clip=true,width=0.7\textwidth]{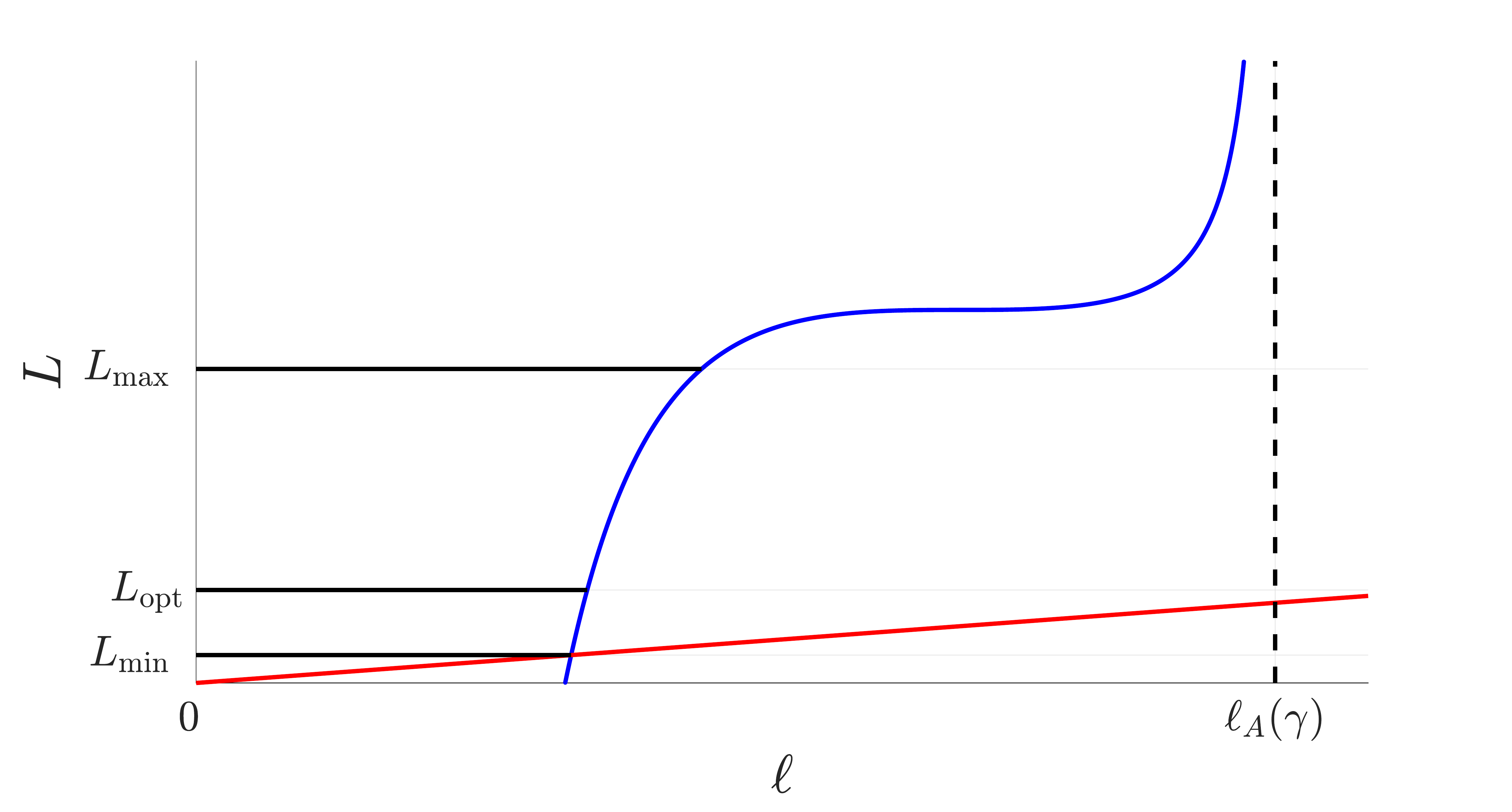}}
\caption{\label{fig:g+p}The function $g \( \ELL \)$ in \eqref{eqn:gL} is plotted in blue, for $\gamma=32$ and $\e=0.05$. The red line is $L = \ELL$. $L_{\min} = 0.224$ as in \eqref{eq:Lmin_defn}, $L_{\text{opt}} = 0.747$ as in \eqref{eqn:Lm}, and $L_{\max} = 2.525$ as in \eqref{eq:Lmax_defn}. The smallest non-zero pole of $g$ occurs at $\ELL = \ELL_A = 0.644$.}
\vspace{-10pt}
\end{figure}

To gain the qualitative insight we need for the three-phase situation, it is useful to consider a single two-phase $B \to A$ interface layer, modelled in the absence of the void. For simplicity, we work on a symmetric interval $y \in \left[ \frac{-L}{2}, \frac{L}{2} \right]$, and seek a continuous profile $\conc$ of the form
\begin{align}\label{eqn:phipw}
\conc(y)=\begin{cases}
-1 & \frac{-L}{2} \leq y\leq y_l\\
\conc_i(y) & y_l < y < y_r \\
1 & y_r \leq y \leq \frac{L}{2},
\end{cases}
\end{align}
with $\left| \conc_i \( y \) \right| < 1$ for $ y_l < y < y_r$. We define $\ELL := y_r - y_l > 0$. In the parlance of our three-phase model, we regard this as a single molecular layer of width $w = L$, having an internal interface layer of width $\ELL$, inside the polymer where $\Vconc = 1$. (Note: setting $\Vconc \equiv 1$ i.e.~$\M = 1$ in \eqref{eq:3phase_1st_variation_dynamics_as_used} causes the three-phase model to break down.) The function $\conc$ is to be a critical point of the two-phase energy functional (cf.~\eqref{eq:ropma} and (5.1) of \cite{choksi_OKFEF_deriv}),
\begin{align}  \label{eqn:eab}
\E_{BA} &= \int_{y_l}^{y_r}{ \dfrac{\e^2}{2} \left| \conc' \right|^2 + \dfrac{1}{4} \( 1 - \conc^2 \) \ud y} + \dfrac{\e \gamma}{2} \int_{-L/2}^{L/2}{ \( \int_{-L/2}^{y}{ \conc \( \eta \) \ud \eta } \)^2 \ud y},
\end{align}
under the constraint $\overline{\conc} = 0$, for a symmetric DBC. The mass constraint implies that
\begin{align}  \label{eqn:icond}
\int_{y_l}^{y_r}{\conc_i \( \eta \) \usd \eta} = y_l+y_r.
\end{align}
Calculating the first variation of $\E_{BA}$ under the given constraint leads to
\begin{subequations}  \label{eqn:ele}
\begin{align}
\e^2 \conc_i''
+ \dfrac{1}{2} \conc_i
+ \e \gamma \int_{-L/2}^{y}{ \( \int_{-L/2}^{\zeta}{ \conc \( \eta \) \usd \eta } \) \ud \zeta }
=\lambda,
\intertext{for $y_l<y<y_r$, with}
\lambda = \dfrac{1}{\ELL} \left[ \frac{y_l+y_r}{2} + \e \gamma \int_{y_l}^{y_r}{ \int_{-L/2}^{y}{ \( \int_{-L/2}^{\zeta}{ \conc \( \eta \) \ud \eta } \) \ud \zeta } \usd y} \right],
\end{align}
\end{subequations}
and 
\begin{align}\label{eqn:cc}
\conc_i \( y_l \) &=-1, & \conc_i \( y_r \) &= 1, & \conc_i ' \( y_l \) &= 0, & \conc_i ' \( y_r \) &= 0.
\end{align}
The first two of these follow from the continuity of $\conc$, which we required \textit{a priori}. This problem is equivalent to 
\begin{equation}\label{eqn:ode}
\e^2 \conc_i'''' + \dfrac{1}{2} \conc_i'' + \e \gamma \conc_i = 0
\end{equation}
together with \eqref{eqn:cc}, and the two additional boundary conditions
\begin{subequations}  \label{eqn:add}
\begin{align}
\e \conc_i ''' \( y_l \) &= \gamma \( y_l + \dfrac{L}{2} \),  \label{eqn:add1}  \\
\e^2 \conc_i'' \( y_l \) &=  \dfrac{1}{2} +
\dfrac{y_r+y_l}{2 \ELL} - \dfrac{\e \gamma}{2 \ELL}\( y_l + \dfrac{L}{2} \) + \dfrac{\e \gamma}{\ELL} \int_{y_l}^{y_r}{ \int_{y_l}^{y} \int_{y_l}^{\zeta}
\conc(\eta) \usd \eta \usd \zeta \usd y}.
\label{eqn:add2}
\end{align}
\end{subequations}
The ODE \eqref{eqn:ode} has the general solution
\begin{subequations}\label{eqn:ogs}
\begin{align}
\conc_i(y)&=
c_1 \sin \( \sqrt{q_h} y \) 
+
c_2 \sin \(\sqrt{q_l} y \)
+
c_3 \cos \(\sqrt{q_h} y \)
+
c_4 \cos \(\sqrt{q_l} y \),\label{eqn:ogs-a} \\ 
q_h &= \dfrac{1}{4 \e^2} \( 1+\sqrt{1-16\gamma\e^3} \),
\quad
q_l = \dfrac{1}{4 \e^2} \( 1-\sqrt{1-16\gamma\e^3} \), \label{eqn:ogs-b}
\end{align}
\end{subequations}
provided that $\gamma\e^3 < \frac{1}{16}$. This bound is not coincidental: a linear stability analysis of the homogeneous state $\conc \equiv 0$ shows that $\gamma\e^3 = \frac{1}{16}$ is a critical value above which the homogeneous, well-mixed state is stable against small perturbations at all wave numbers. The constants $c_1$, $c_2$, $c_3$ and $c_4$ along with $y_l$ and $y_r$ are found by substituting \eqref{eqn:ogs} into \eqref{eqn:cc} and \eqref{eqn:add}. As shown in appendix \ref{appendix:symmetry}, if $c_B$ defined there in \eqref{eqn:cB} is non-zero, then $y_r=-y_l = \frac{\ELL}{2}$, and $c_3=0=c_4$, and we are left with
\begin{subequations}\label{eqn:3sys}
\begin{align}
c_1 \sin{\( \sqrt{q_h} \frac{\ELL}{2} \)} + c_2 \sin{\(\sqrt{q_l} \frac{\ELL}{2} \)} &= 1,  \label{eqn:3sysa} \\
c_1 \sqrt{q_h} \cos{\( \sqrt{q_h} \frac{\ELL}{2} \)} + c_2 \sqrt{q_l} \cos{\( \sqrt{q_l} \frac{\ELL}{2} \)} &= 0, \label{eqn:3sysb} \\
\dfrac{c_1}{\sqrt{q_h}} \cos{\( \sqrt{q_h} \frac{\ELL}{2} \)} + \frac{c_2}{\sqrt{q_l}} \cos{\( \sqrt{q_l} \frac{\ELL}{2} \)} &= \dfrac{L - \ELL}{2},  \label{eqn:3sysc}
\end{align}
\end{subequations}
for the three unknowns $c_1$, $c_2$ and $\ELL$. This is the generic case, so we do not pursue the alternative $c_B=0$ any further. If, on the other hand,
\begin{align}
c_A \( \ELL,\gamma \) := \sqrt{q_h} \cos{\( \sqrt{q_h} \frac{\ELL}{2} \)} \sin{\( \sqrt{q_l} \frac{\ELL}{2} \)} -  
\sqrt{q_l} \cos{\( \sqrt{q_l} \frac{\ELL}{2} \)} \sin{\( \sqrt{q_h} \frac{\ELL}{2} \)} &= 0,
\end{align}
then the system \eqref{eqn:3sys} does not have a solution. Otherwise, we can
solve \eqref{eqn:3sysa} and \eqref{eqn:3sysb} for $c_1$ and $c_2$ and insert them 
into \eqref{eqn:3sysc} to give (cf.~Fig.~\ref{fig:g+p})
\begin{equation}\label{eqn:gL}
g \( \ELL; \gamma \) := \dfrac{ 2 \( q_h - q_l \)}{c_A \sqrt{q_h q_l}} \cos{\( \sqrt{q_h} \frac{\ELL}{2} \)} \cos{\( \sqrt{q_l} \frac{\ELL}{2} \)} + \ELL = L.
\end{equation}

We wish to consider the solutions $\ELL > 0$ of \eqref{eqn:gL} as we vary $\gamma$, $\e$ and $L$. Notice that we can reduce the number of paramters to two by rescaling  $\hat y=y/\e$, $\hat L= L/\e$, and letting $\rho=16\gamma\e^3$.  Instead of explicitly carrying out this rescaling however, we fix $\gamma$ and vary $\e$, but refer to $\rho$ where this is useful. For the purpose of presenting plots, we set $\gamma=32$, to be consistent with our choices in section~\ref{sec:num}. 

We first consider $\e \to 0$ which, for fixed $\gamma$, is equivalent to $\rho \ll 1$. The function $g$ shown in Fig.~\ref{fig:g+p} has $\e=0.05$ and $\gamma=32$ viz.~$\rho=0.064$.  For $ \ELL > 0$, $g\( \ell; \gamma \)$ has poles where $c_A \( \ELL, \gamma \)=0$. Between these poles, $g$ increases monotonically with $\ELL$, and thus \eqref{eqn:gL} has one solution $\ELL$ between each pair of consecutive poles, for a given $L>0$.  We focus on solutions corresponding to the thinnest interface layers, for which $0 < \ELL < \ELL_A \( \gamma \)$, where $\ELL_A \( \gamma \)$ is the first non-zero pole of $g$ (for a given $\gamma$).

Since the \textit{ansatz} \eqref{eqn:phipw} implicitly assumes that $L \geq \ELL$, we are only interested in solutions of \eqref{eqn:gL} where $g> \ELL$. This holds if and only if $L > L_{\min}$ 
(or equivalently, $\ELL > L_{\text{min}}$), where 
\begin{align}  \label{eq:Lmin_defn}
L_{\min} := \dfrac{\pi}{\sqrt{q_h}} \sim \sqrt{2} \pi \e.
\end{align}
On the other hand, we find that $\left| \conc \right| > 1$ near $y=y_l$ and $y=y_r$ if $L$ or $\ELL$ are too large. To avoid such profiles it was sufficient to have $\ELL < \ELL_A/2$, but this effectively imposes the upper bound $L < L_{\max} := g \( \ell_A/2 \)$, too. In the sharp interface limit, it turns out that 
\begin{align}  \label{eq:Lmax_defn}
L_{\max} \sim \frac{0.202}{\gamma\e^2}.
\end{align}

The optimal molecular layer width $L=L_{\text{opt}}$ is defined as that $L$ for which the energy density $\E_{BA}/L$ is minimal. To find this value in the sharp interface limit, we
first determine an asymptotic approximation for the solution of \eqref{eqn:3sys}, as $\e \to 0$, with $\gamma$ and $L$ fixed. We obtain the expansion for $\ELL$ from \eqref{eqn:gL}, and then use it to expand the solutions for $c_1$ and $c_2$ of \eqref{eqn:3sysa} and \eqref{eqn:3sysb}; we find that
\begin{align}  \label{eqn:exp}
\ELL &\sim \sqrt 2 \pi\e + 4\gamma L\e^3, & 
c_1 &\sim 1, & &\text{and} & 
c_2 &\sim \dfrac{\sqrt{2}}{2} \gamma^{1/2} L \e^{1/2}.
\end{align}
Using this and \eqref{eqn:ogs} in the energy \eqref{eqn:eab} gives, after expanding,
\begin{align}  \label{eqn:EABrho}
\E_{BA} &\sim \( \frac{\gamma L^3}{24} + \dfrac{\pi}{2 \sqrt{2}} \) \e.
\end{align} 
The energy density $\E_{BA}/L$ is thus minimized when
\begin{align}  \label{eqn:Lm}
L=L_{\text{opt}} \sim \dfrac{\( 3 \sqrt{2} \pi \)^{1/3}}{\gamma^{1/3}} = w_m,
\end{align} 
in agreement with \eqref{eqn:wmin-si}. This value satisfies $L_{\text{min}}<L_{\text{opt}}<L_{\text{max}}$ as $\e\to 0$, as required for a useful model.

On the other hand, our numerical parameter study revealed that $L_{\text{opt}}$ tends to $L_{\text{min}}$ as $\rho\to 1$. This is in agreement with our earlier observation that $\rho=1$ is the threshold for the linear stability of a homogeneous mixture of the $A$ and $B$ species (i.e.~$\conc \equiv 0$), and implies that we need to observe the upper bound
\begin{equation}\label{eqn:epsbd}
\e < \frac{1}{2 \( 2 \gamma \)^{1/3}}.
\end{equation}
Since we also wish to represent monolayers which are thinner (for example in a 1D setting, where an integer number of monolayers has to fit into the domain), $\e$ needs to be somewhat smaller than this. A simple approach is to choose an upper bound for $\e$ that is half of the bound in \eqref{eqn:epsbd}. This has a manageable impact on grid resolution, but guarantees the bound $\rho\leq \frac{1}{8}$, i.e. $\rho \ll 1$, so that \eqref{eqn:exp}--\eqref{eqn:EABrho} are valid, and the difficulties with having $L_{\text{opt}}$ too close to $L_{\text{min}}$ are avoided.

\section{Numerical simulations}  \label{sec:num}

We now present our numerical solutions of the time-dependent, regularised problem \eqref{eqn:timedep}, \eqref{eq:3phase_1st_variation_dynamics_as_used}.  We used a $\mathbb{P}_1$ finite element numerical discretisation of space (implemented in \textsc{FEniCS} \cite{AlnaesBlechta2015a}), and a one-step Euler discretisation of time. We used the semi-smooth Newton (SSN) method (generalising the method from \cite{bosch2014fast}) to resolve the non-linearity at each time-step. For simplicity, we used a fixed spatial grid and and variable time-stepping. We coded the time integration to recursively retry any time-step $\tau$ for which SSN did not converge in ten iterations, at $\tau/2$. We denote our simulation time horizon $\( 0, T \right]$, and increased $T$ in each experiment until the decay in the energy was essentially zero, and the profiles could be considered stationary. We worked spatially in 1D and 2D on the domains $\( 0, L \)$ and $\( 0, L_x \) \times \( 0, L \)$; values for $L$ and $L_x$ can be read off the figures. 

\begin{figure}[tp]
\begin{center}
  \subfloat[$\conc$ end-state profiles]{\label{fig:1D_phi_profiles}\includegraphics[bb=0 0 524 385, clip=true,width=0.45\textwidth]{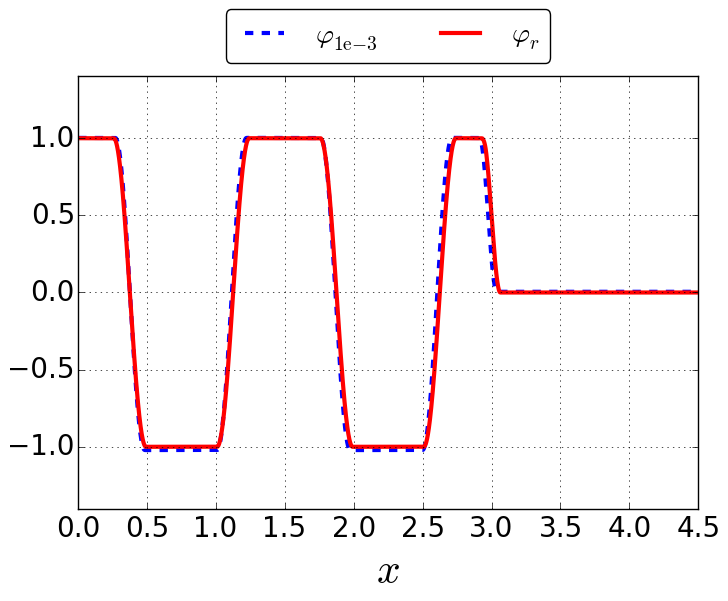}}
  \subfloat[$\Vconc$ end-state profiles]{\label{fig:1D_psi_profiles}\includegraphics[bb=0 0 524 385, clip=true,width=0.45\textwidth]{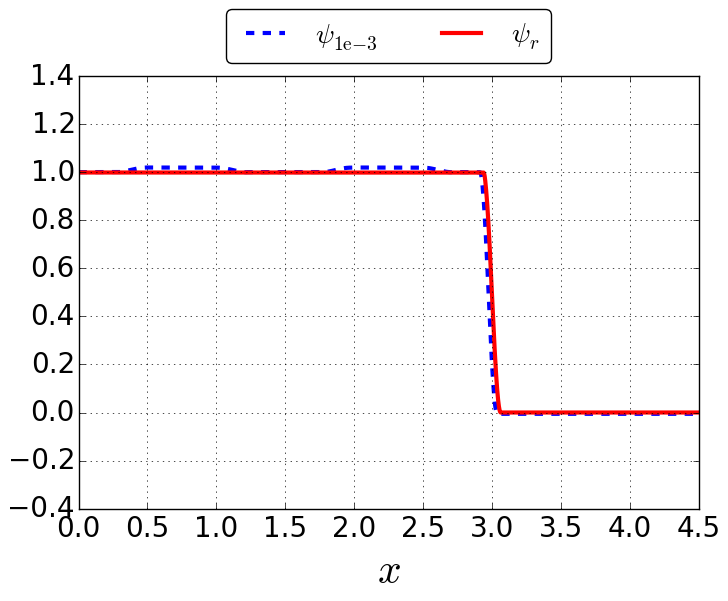}}
\caption{\label{fig:res:1d}One-dimensional steady-state solutions obtained by solving the time-dependent model \eqref{eq:3phase_1st_variation_dynamics_as_used} with initial condition \eqref{eqn:run1d:ic}, the parameters \eqref{eqn:parms:def}, \eqref{1dparm} with $\mu=10^{-3}$ (dashed blue lines) and $\mu=10^{-8}$ (solid red lines).}
\end{center}
\vspace{-10pt}
\end{figure}

We worked with two rectangular geometries, similar to those used in the first numerical SCFT experiments in \cite{stasiak2012step}. In the first, we considered a stationary thin film having a step from one to three monolayers (the $L_{1}^{||}$--$L_{3}^{||}$ configuration of \cite{stasiak2012step}) on a 2D domain with $L_x = 16$ and $L = 3$. Our second 2D geometry was a rectangle with $L_x = 8$ and $L = 4.5$, on which we explored variants of the $L_{2}^{||}$--$L_{4}^{||}$ configuration of \cite{stasiak2012step}. Unless otherwise stated, we set the polymer volume fraction to $\mathcal{M}=0.5$. In both geometries, this fixed the optimal (vertical) width of the monolayers in equilibrium at $w_m=L_{\text{opt}}=0.75$, from which we could infer $\gamma$ via \eqref{eqn:Lm} (or \eqref{eqn:wmin-si}) and then choose an $\e$ that was lower than half of the bound in \eqref{eqn:epsbd}. We selected
\begin{align}  \label{eqn:parms:def} 
\gamma &= 32, & \e &= 0.05, & \K &= 1,
\end{align}
where the latter results in similar thicknesses for the $AB$ and the polymer--void interface layers, if $\chi^{AV}$ is close to 1. 
\subsubsection*{1D results}
We conducted a series of 1D experiments on $\( 0, L \)$, $L=4.5$, representing a vertical section through our second 2D geometry of interest (see Fig.~\ref{fig:res:1d}). We wished to understand the effect of $\mu$ so as to make a good choice for this parameter. We used as an initial condition the interpolant of
\begin{subequations}  \label{eqn:run1d:ic}
\begin{align}
\conc \left( x \right) &= 0.5 + \sum_{i=1}^{4}{(-1)^i H_\delta \left( x, \dfrac{3 (i-0.5)}{4}, 0.07 \right)} - 0.5 H_\delta \left( x, 3, 0.04 \right) \\
\Vconc \left( x \right) &= 0.5 \left( 1 - H_\delta \left( x, 3, 0.04 \right) \right),
\end{align}
\end{subequations}
where
\begin{align*}
H_\delta \left( x, X, \d \right) = 
\begin{cases}
-1 & \dfrac{x - X}{\d} < \dfrac{- \pi}{2} \\
\sin{ \left( \dfrac{x - X}{\d} \right)} & \left| \dfrac{x - X}{\d} \right| \leq \dfrac{\pi}{2} \\
+1 & \dfrac{x - X}{\d} > \dfrac{\pi}{2}, \end{cases}
\end{align*}
comprising four zero-mass internal interface layers, evenly-spaced between $x=0$ and $x=3$, viz.~the midpoint of the single polymer--void interface layer. We used \eqref{eqn:parms:def}, along with
\begin{align}  \label{1dparm}
\chi^{AV}&=1.5, & \chi^{\D}&=18.5, & \mathcal{M} &= \frac{2}{3},
\end{align}
and a range of $\mu$-values.  The results for two different values of $\mu$ are shown in Fig.~\ref{fig:res:1d}. Clearly, $(\conc, \Vconc)$ is close to $(1,1)$ (pure $A$), $(-1,1)$ (pure $B$), $(0,0)$ (pure $V$), except in the interface layers. 

On closer inspection of Fig.~\ref{fig:res:1d}, it is evident that for $\mu=10^{-3}$, $\conc$ and $\Vconc$ visibly exceed the range of values permitted by the obstacle potential. In fact, for $\mu \in \left\{ 10^{-3}, 10^{-4}, \ldots, 10^{-8} \right\}$, we found that
\[
\max \left\{ \max \left( \Vconc - 1 \right)_+, \max \left( \conc - \Vconc \right)_+, - \min \left( \conc + \Vconc \right)_- \right\}
\]
decreases asymptotically linearly as $\mu \to 0$. At $\mu=10^{-8}$ this discrepancy was $2 \times 10^{-7}$. Besides altering the plateau values between the interface layers, Fig.~\ref{fig:res:1d} also suggests that the regularisation shifts the layers horizontally by a similar amount.

For the energy of the stationary state, the convergence is also linear in $\mu$.  At $\mu=10^{-8}$, we get $\E_\mu=0.37745$, and only the last digit changes if $\mu$ is increased by a factor of $10$; hence we take $\E=0.3775$ as the converged value. In contrast, $\E_{\mu} = 0.016$ and $0.342$ if $\mu=10^{-3}$ and $\mu=10^{-4}$. This is consistent with a linear convergence in $\mu$, but also suggests that for $\E_{\mu}$ to be within 10\% of $\E$, we must have $\mu\leq 10^{-4}$. The sharp interface energy estimate from \eqref{eq:i0nm}, with $N=4$, $N_{AV}=1$, $N_{BV}=0$, $L=4.5$, is $I_0=0.383$, which is close to $\E$. That said, Fig.~\ref{fig:res:1d} suggests that the profile shape is qualitatively correct, even with the relatively large $\mu = 10^{-3}$. On the other hand, informal experimentation in 2D suggested that average SSN iteration counts grow as $- \log \mu$. To balance the desire for accuracy against our time and compute budget, we thus selected $\mu=10^{-6}$ as the default for our 2D experiments.

\begin{figure}[tp]
\begin{center}
  \subfloat[Shared initial condition: $t = 0$]{\label{fig:set1_common_IC}\includegraphics[clip=true,width=0.31\textwidth]{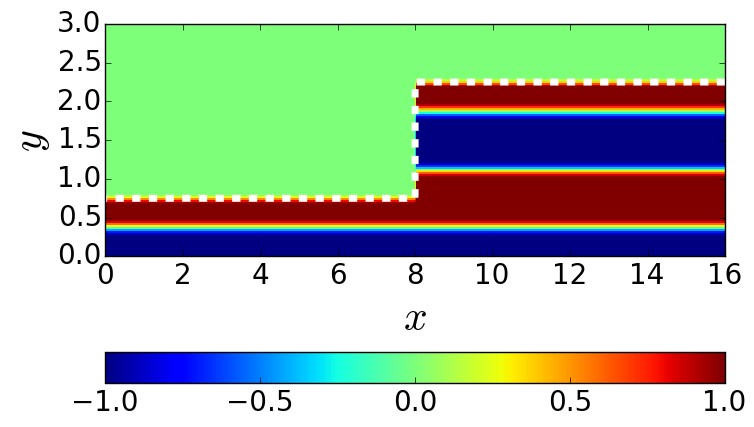}} \hspace{5pt}
  \subfloat[The open configuration: $\chi^{\D} = 1.5$; \eqref{parm:run2top}. $\E_{\mu} = 3.441$]{\label{fig:2D_open_phi_end_state}\includegraphics[clip=true,width=0.31\textwidth]{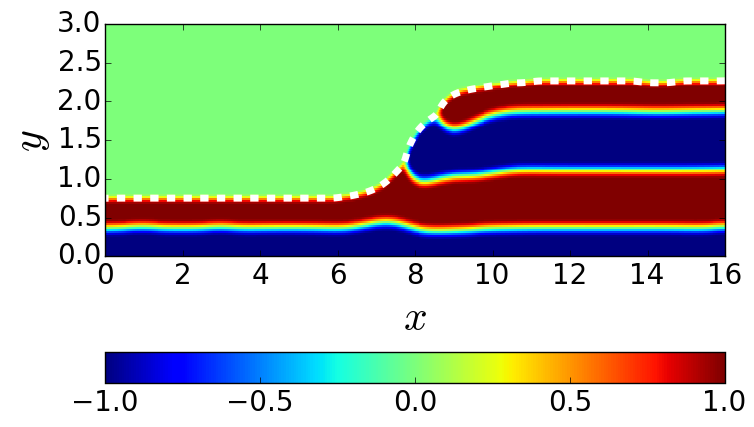}}  \hspace{5pt}
  \subfloat[The closed configuration: $\chi^{\D} = 18.5$; \eqref{parm:run2bottom}. $\E_{\mu} = 3.461$]{\label{fig:2D_closed_phi_end_state}\includegraphics[clip=true,width=0.31\textwidth]{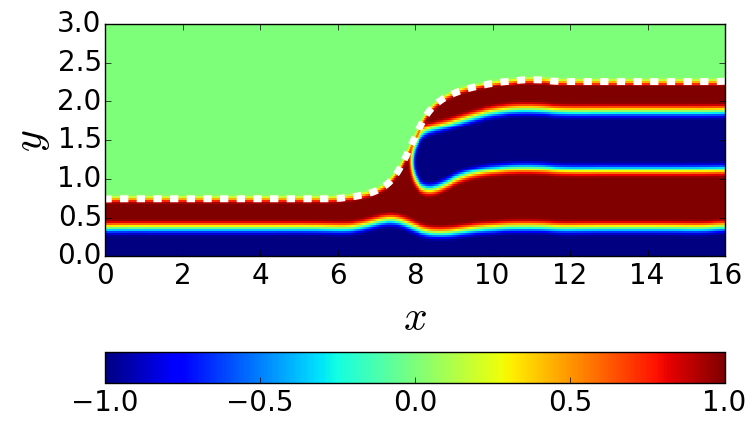}} 
\caption{\label{fig:Neq2-oc}Plots of $\conc$ for the first and second 2D simulations; the colour map and the dashed white line are explained in the text. The parameters in \eqref{eqn:parms:def} were used in both experiments.}
\end{center}
\vspace{-12pt}
\end{figure}

\subsubsection*{2D results}
We first explain the colour map used in Fig.~\subref*{fig:set1_common_IC}, and all subsequent 2D figures. Light green represents $\conc=0$, dark red $\conc= 1$, and dark blue $\conc= - 1$. $\Vconc$ has a simple topology so we do not introduce a separate colour scheme. Instead, we include the dashed white line marking the contour line $\Vconc=\frac{1}{2}$, to indicate the middle of the polymer--void interface layer. 
The light green area above the dashed white line represents the region with the (nearly) pure $V$ (void) species, $\Vconc \approx 0$, while the region below is occupied by the polymer, $\Vconc \approx 1$. Red and blue regions indicate the (nearly pure) $A$ ($\conc \approx 1$) and $B$ ($\conc \approx -1$) microdomains, with other colours indicating the layers between them. 

Where relevant, we compared our 2D end-state energies with the corresponding minimum flat film energy, defined as the product of $L_x$ and the minimum energy of a vertical 1D section configuration (obtained by simulation) with $\mathcal{M} = 0.5$. We then compared these with the sharp interface estimates from \eqref{eq:i0nm} and \eqref{eq:generic_2D_stepped_energy_est} (with $\xi = \frac{1}{2}$) using, for the flat alternative and the $L_{N_1}^{||}$--$L_{N_2}^{||}$ stepped configuration,
\begin{align}  \label{eq:2D_sharp_interface_energy_estimates}
\E_{\text{flat}} \( N \) &= I_0 \( N \) \times \e L_x
& 
\E_{\text{step}} \( N_1, N_2 \) &= \frac{I_0 \( N_1, \M_1 \) + I_0 \( N_2, \M_2 \)}{2} \times \e L_x,
\end{align}
where we have suppressed $\M = \frac{1}{2}$ on the left, and $\frac{1}{2} \( \M_1 + \M_2 \) = \M = \frac{1}{2}$.

For our first pair of 2D experiments, we used the initial configuration in Fig.~\subref*{fig:set1_common_IC}, in which the domain was split into left and right halves. On the left, $0<x<L_x/2$, we used a configuration with a $B$-microdomain at the substrate ($y=0$) of width $w_m/2$, below an $A$-microdomain of the same width (i.e.~a one layer configuration).  For $L_x/2<x<L_x$, we had four microdomains $BAABBA$ (three monolayers), of width $w_m/2$, $w_m$, $w_m$ and $w_m/2$. Because of our spatial discretisation, $\overline{\conc_0}$ was not exactly zero (this was true for all of our 2D experiments). Thanks to the mass control property of \eqref{eq:3phase_1st_variation_dynamics_as_used} however, this was not a serious issue. The space above the polymer material was occupied by the void $V$, with $\mathcal{M} = 0.5$.  Since $w_m$ is the optimal monolayer width in the sharp interface limit, we expected the initial configuration to be close to the final equilibrium. Indeed, the expression \eqref{eqn:steploc} predicts that for $N_1=1$, $N_2=3$, and $\M L =\( N_1 + N_2 \) w_m/2$, the step should remain very close to $x=\xi L_x=L_x/2$.  In the numerical simulation, in which we set
\begin{align}  \label{parm:run2top}
\chi^{AV} &= 1.5, & \chi^{\D} &= 1.5,
\end{align}
there was initially a fairly rapid evolution as the side of the step relaxed and the polymer--void interface layer adjusted. At $T = 202$, the solution shown in Fig.~\subref*{fig:2D_open_phi_end_state} was essentially stationary. The step was still centered around $x=L_x/2$.  The topology at the surface was largely unchanged, with the $A$ phase occupying most of the interface layer with the void, except for a small portion in the step transition near $x=8$. We also observed two triple points where all three phases met. The flat configuration with lowest energy was the two-layer $BAABV$ with $\E_{\mu} = 3.619$ (which exceeds that noted below Fig.~\subref*{fig:2D_open_phi_end_state}). The sharp interface energy estimates \eqref{eq:2D_sharp_interface_energy_estimates} were $\E_{\text{flat}} \( 1 \) = 5.258$, $\E_{\text{flat}} \( 2 \) = 3.765$, $\E_{\text{flat}} \( 3 \) = 3.835$ and $\E_{\text{step}} \( 1, 3 \) = 3.447$ (smallest), in good agreement with our numerical results.

\begin{figure}[tp]
\begin{center}
  \subfloat[$t = 0$]{\label{fig:c0_initial_condition}\includegraphics[clip=true,width=0.45\textwidth]{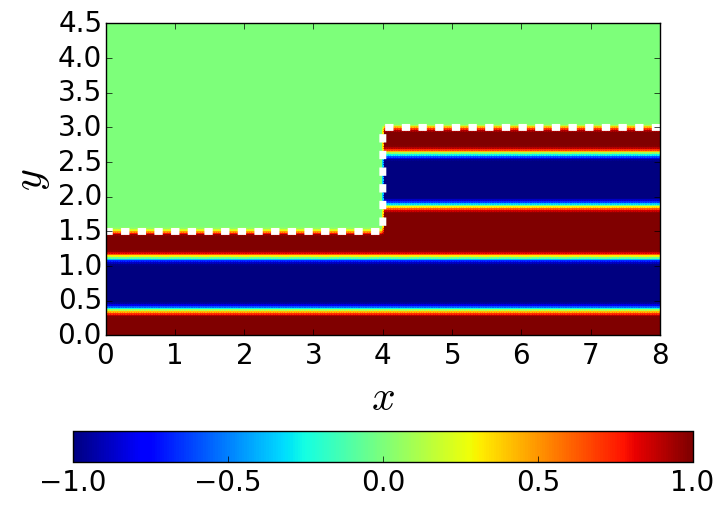}} \hspace{10pt}
  \subfloat[The $C_0$ closed configuration. $\chi^{\D} = 18.5$ as in \eqref{parm:run2bottom}. $\E_{\mu} = 2.408$]{\label{fig:c0_end_state}\includegraphics[clip=true,width=0.45\textwidth]{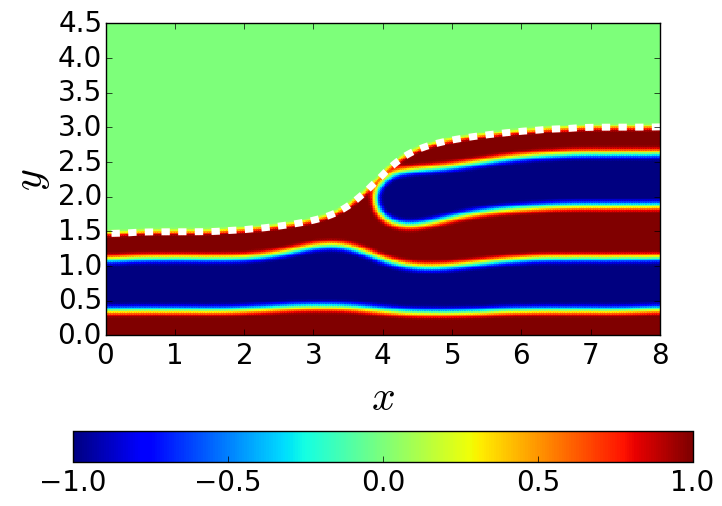}} \\ \vspace{-10pt}
  \subfloat[$t = 0$]{\label{fig:c1_initial_condition}\includegraphics[clip=true,width=0.45\textwidth]{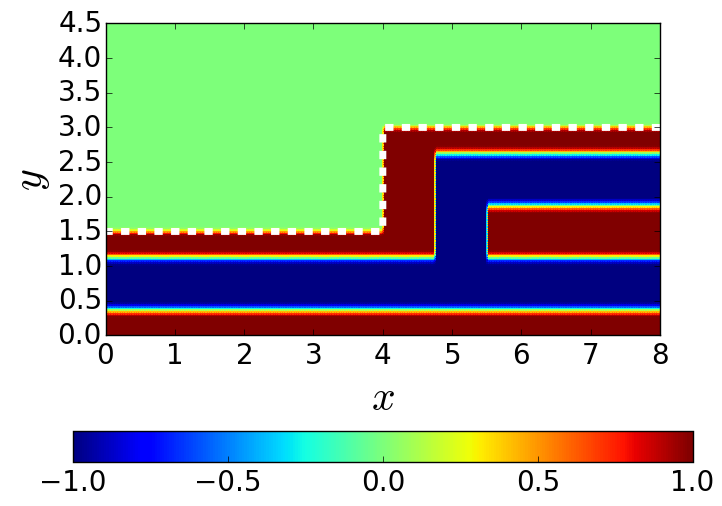}} \hspace{10pt}
  \subfloat[The $C_1$ closed configuration. $\chi^{\D} = 18.5$ as in \eqref{parm:run2bottom}. $\E_{\mu} = 2.411$]{\label{fig:c1_end_state}\includegraphics[clip=true,width=0.45\textwidth]{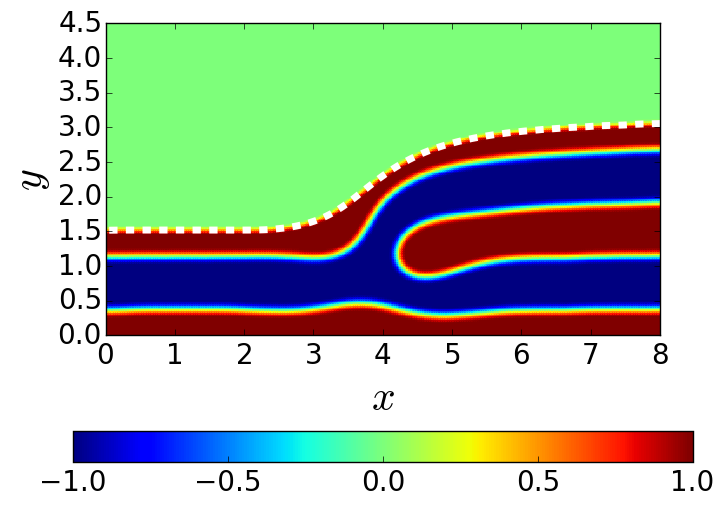}}
\caption{\label{fig:l2l4-c0c1}Initial conditions and end-states. These are $L_{2}^{||}$--$L_{4}^{||}$ configurations in \cite{stasiak2012step}.}
\end{center}
\vspace{-12pt}
\end{figure}    

If the surface tension of one of the interface layers, say ${BV}$, exceeds the sum of the other two, $\sigma_{BV}>\sigma_{AB}+\sigma_{AV}$, then it is energetically favourable to remove the triple-phase contact line by eliminating the interface layer between $B$ and $V$ completely. This is the situation in the bottom row of Fig.~\ref{fig:Neq2-oc}, in which we increased $\chi^{\D}$, setting
\begin{align}\label{parm:run2bottom}
\chi^{AV} &= 1.5, 
& 
\chi^{\D} &= 18.5.
\end{align}
The other parameters and initial conditions were unchanged. From \eqref{eqn:surface-tension-values}, the surface tensions are $\sigma_{AB}=1.11$, $\sigma_{AV}=0.96$, and $\sigma_{BV}=3.51$.  Observe that the triple points have moved towards each other, leading to the closed configuration in Fig.~\subref*{fig:2D_closed_phi_end_state}, recorded at $T = 102$, in which $B$ and $V$ do not share an interface layer (cf.~the open configuration of Fig.~\subref*{fig:2D_open_phi_end_state}). 

For this much higher value of $\chi^{\D}$, the simulated flat configuration with the lowest energy was the three-layer $BAABBAV$ (notice the $AV$ interface layer) with $\E_{\mu} = 3.792$ (exceeding that noted below Fig.~\subref*{fig:2D_closed_phi_end_state}). The sharp interface energy estimates \eqref{eq:2D_sharp_interface_energy_estimates} were the same as those above, except for $\E_{\text{flat}} \( 2 \) = 5.487$ (because of the $BV$ interface layer). The value of $w_m$ in \eqref{eqn:wmin-si} corresponded to the flat two-layer configuration which had much higher energy; see the discussion immediately preceding section \ref{subsec:finite}.

In our second pair of 2D experiments, summarised in Fig.~\ref{fig:l2l4-c0c1}, we used the same parameters \eqref{eqn:parms:def} and \eqref{parm:run2bottom} as in Fig.~\subref*{fig:set1_common_IC} and \subref{fig:2D_closed_phi_end_state}, but used initial conditions having two and four monolayers to the left and right of the step, and an $A$-microdomain at the substrate. In the first experiment, the top $B$-microdomain to the right of the step (the second blue layer in Fig.~\subref*{fig:c0_initial_condition}) was a semi-layer; in the second (Fig.~\subref*{fig:c1_initial_condition}), the second $A$-microdomain was a truncated layer. Using notation from \cite{stasiak2012step}, we call these the $C_0$ and $C_1$ topologies. To keep the polymer volume fraction constant at $\mathcal{M}=0.5$ (for the same optimal layer width as our previous experiments), the vertical extent of the domain was increased to $L=4.5$. The lateral extent was decreased to limit computational cost.  These very different topologies were preserved as the systems evolved, giving the results in Fig's~\subref*{fig:c0_end_state} (recorded at $T = 133$) and \subref{fig:c1_end_state} (recorded at $T = 193$). At the time for which the results are shown, the energy decay had essentially stopped, suggesting that both configurations were very close to stable steady states, that is, local energetic minima. The total energies of these states are very similar to those noted in \cite{stasiak2012step}. However, while those authors recorded a lower energy for the $C_1$ than for the $C_0$ topology, our energy values were too close for a reliable comparison, given our spatial grid resolution (about 40 nodes per unit length in each direction).

The corresponding flat configuration with the lowest energy, for both of these scenarios, was the four-layer $ABBAABBAV$ configuration with $\E_{\mu} = 2.509$, which exceeds both of the values recorded in Fig's \subref*{fig:c0_end_state}, \subref{fig:c1_end_state}. The sharp interface estimates \eqref{eq:2D_sharp_interface_energy_estimates} were $\E_{\text{flat}} \( 2 \) = 2.792$, $\E_{\text{flat}} \( 3 \) = 3.413$, $\E_{\text{flat}} \( 4 \) = 2.542$ and $\E_{\text{step}} \( 2, 4 \) = 2.393$ (smallest). Our sharp interface and numerical results were thus in qualitative agreement.

\begin{figure}[tp]
\begin{center}
  \subfloat[$t \approx 30 $]{\label{fig:phi_end_state_exp_run_2642}\includegraphics[clip=true,width=0.45\textwidth]{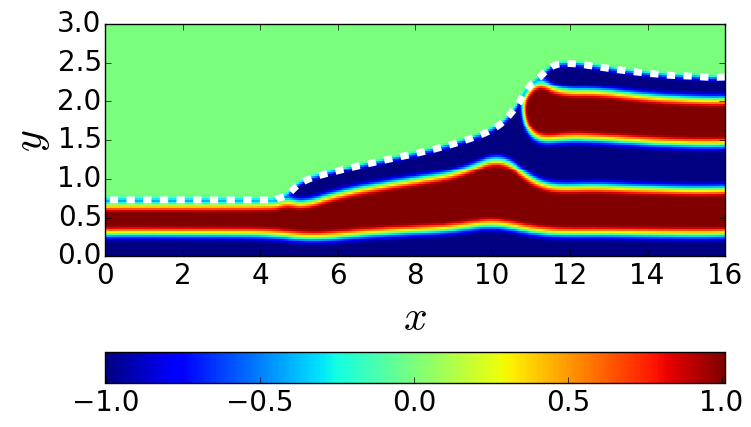}} 
  \subfloat[$t \approx 90$]{\label{fig:phi_end_state_exp_run_2708}\includegraphics[clip=true,width=0.45\textwidth]{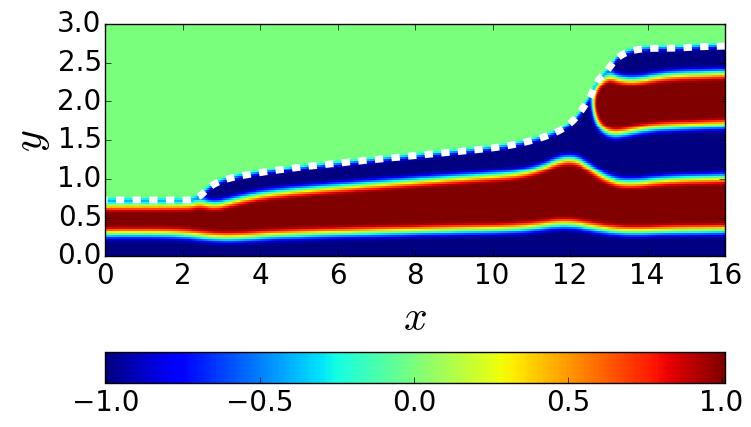}} \\ \vspace{-10pt}
  \subfloat[$t \approx 280$]{\label{fig:phi_end_state_exp_run_2726}\includegraphics[clip=true,width=0.45\textwidth]{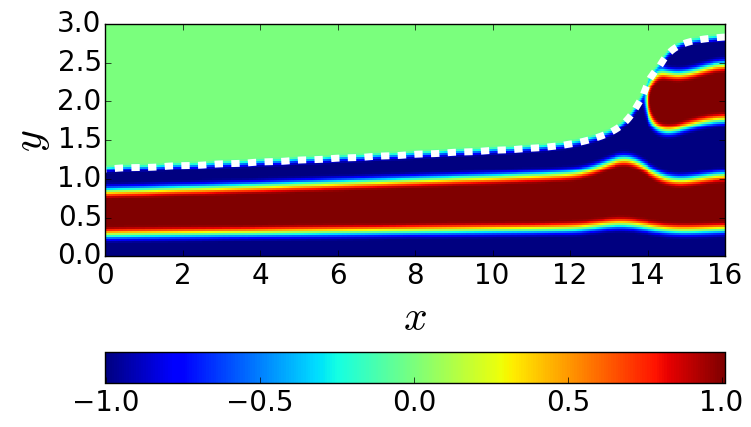}}
  \subfloat[$T \approx 1530$]{\label{fig:phi_end_state_exp_run_2762}\includegraphics[clip=true,width=0.45\textwidth]{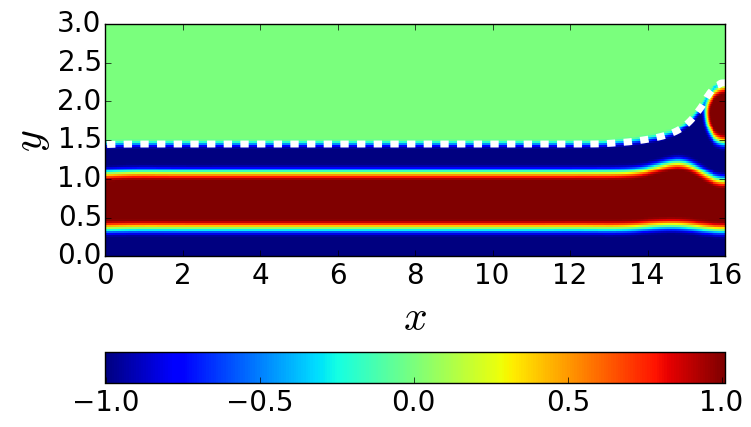}} 
\caption{\label{fig:inv}Results at the times indicated for parameters \eqref{eqn:parms:def} and \eqref{eqn:parms:invert}, for which a $BV$ polymer--void interface layer is energetically preferred to an $AV$ interface layer.}
\end{center}
\vspace{-12pt}
\end{figure}

Our final 2D experiment on $\( 0, 16 \) \times \( 0, 3 \)$ is summarised in Fig.~\ref{fig:inv}. We started with the configuration depicted in Fig.~\subref*{fig:set1_common_IC} and the parameters \eqref{eqn:parms:def}, but inverted the preferences between the void and the polymer species by setting 
\begin{align}\label{eqn:parms:invert}
\chi^{AV}&=10, & \chi^{\D}&= -8.5.
\end{align}
The $BV$ interface layer was thus preferred to the $AV$ layer. As a result, the blue semilayer spread over the $AV$ interface layer to the left and right. At the same time, the top red semilayer retracted as the $A$-species diffused across the blue $B$-microdomain into the other $A$-microdomain, until all that remained was a small semi-circle at $x=L_x$. At this stage, the evolution effectively stopped, suggesting that a solution close to a (stable) stationary state had been reached. The persistence of the residual semi-circle was counter-intuitive. For this much longer running simulation, we used $\mu = 10^{-3}$; as noted previously, we nevertheless expect the configurations depicted in Fig.~\ref{fig:inv} to be qualitatively correct.

\section{Conclusions and outlook}  \label{sec:conclusions}

We have used an Ohta--Kawasaki DFT model with an obstacle potential for the homogeneous free energy to investigate the formation of steps in thin 2D films of a symmetric DBC. The free surface was captured by treating the void as a third phase. These films typically form layers that are aligned with the substrate, with the species at the substrate and the free surface determined by energetic preferences. In this paper, we studied such solutions exclusively, and did not consider other orientations or 3D geometries \cite{PhysRevLett.97.204502, mcgraw2011dynamics}.

An energetically stable, flat film requires the film thickness to be a multiple of $w_m$, the thickness of the monolayer with the smallest energy density.  If $V_p/L_x$ is not an integer multiple of $w_m$, where $V_p$ is the total area occupied by the copolymer, it is possible for stepped films to have lower energies than flat films with a non-optimal monolayer width. In the sharp interface limit, $\e \to 0$ with $\gamma$ fixed, the optimal step location is determined by $w_m$ and mass conservation, according to the closed analytic expression \eqref{eqn:steploc}. Numerical simulations of the ternary Ohta--Kawasaki model confirmed these estimates, and revealed further details of the step between the two film thicknesses. They revealed that the distinction between open and closed configurations depends on the wettability properties of the two species at the free surface. We also explored model properties for a larger number of monolayers, which allow for a larger variety of configurations, by simulating the $C_0$ and $C_1$ closed configurations. These are long-time, persistent solutions, in which the $C_1$ configuration is energetically favourable. Hence, the model and its sharp interface limit captured the essential features of step formation observed experimentally, and in SCFT calculations \cite{stasiak2012step}.

The use of a time-dependent model allowed us to study film dynamics under the assumption that bulk diffusion is the dominant migration mechanism for the polymer species. We studied an example in which the top layer in the initial step configuration was the less preferred layer, and observed the step retract as the preferred species wet the free surface. This eventually led to a nearly flat film that was only perturbed near the boundary of the domain.  Such a configuration may be amenable to experimental investigation using the technique in \cite{Maher2016a}, whereby the film is covered by an additional top layer of a specifically chosen homopolymer.  

The use of an obstacle potential has the advantage that the phase-field variables are uniform, and assume the value for a pure species outside the boundary layer. This allows for exact, explicit representations, and facilitates the analysis of layered patterns, similar to the scalar Cahn--Hilliard equation with obstacle potential \cite{BloweE91}. Suitable interface layer solutions  exist if $\e$ is below a certain bound proportional to $\gamma^{-1/3}$, which is satisfied for well-segregated copolymer films where the SCFT becomes numerically challenging \cite{stasiak2012step}. Phase-field approaches such as the obstacle Ohta--Kawasaki model used here are thus attractive
to SCFT approaches for these situations, and provide and interesting perspective for 3D simulations of copolymer free films, where a rich bulk geometry interacts with the free surface to form terraces. These have been investigated using dynamic variants of the SCFT, but resolving the structures of interest on sufficiently large domains is computationally demanding, and may rekindle interest in phase-field approaches for such large-scale computations  \cite{lyakhova2006dynamics}.

\subsubsection*{Acknowledgements}
The first author thanks Patrick Farrell (Mathematical Institute, Oxford University) and Chris Richardson (BP Institute, Cambridge University) for help with \textsc{FEniCS}. Quentin Parsons was funded by an Oxford University Computer Science EPSRC DTP grant.

\begin{appendix}

\section{Symmetry of the solution in the interface layer}\label{appendix:symmetry}

In this appendix we show that it sufficient to consider only solutions \eqref{eqn:ogs} with $c_3=0=c_4$ and $y_l=-y_r$. We start from a modified version of \eqref{eqn:ogs-a}, 
\begin{gather}
\conc_i(y)=
c_1\sin\left(\sqrt{q_h} \( y - \overline{y} \) \right) 
+
c_2\sin\left(\sqrt{q_l} \( y - \overline{y} \) \right)
+
c_3\cos\left(\sqrt{q_h} \( y - \overline{y} \) \right)
\notag \\
+
c_4\cos\left(\sqrt{q_l} \( y - \overline{y} \) \right)
\label{eqn:dans}
\end{gather}
with $\overline{y}=\( y_l+y_r \)/2$. Plugging this into \eqref{eqn:cc} and \eqref{eqn:add} with $\ELL=y_r-y_l$, yields the problem $A \boldsymbol{c} = \boldsymbol{m}$ with
\begin{subequations}  \label{eqn:4x4}   
\begin{align}
\boldsymbol{c} &= \( c_1, c_2, c_3, c_4 \)^{\top}, & \boldsymbol{m} &= \( 1, -1, 0, 0 \)^{\top},
\end{align}
and
\begin{align}
A = \( 
\begin{array}{cccc}
\sin\( \sqrt{q_h} \frac{\ELL}{2} \) & \sin\( \sqrt{q_l} \frac{\ELL}{2} \) & \cos\( \sqrt{q_h} \frac{\ELL}{2} \) & \cos\( \sqrt{q_l} \frac{\ELL}{2} \) \\
-\sin\( \sqrt{q_h} \frac{\ELL}{2} \) & - \sin\( \sqrt{q_l} \frac{\ELL}{2} \) & \cos\( \sqrt{q_h} \frac{\ELL}{2} \) & \cos\( \sqrt{q_l} \frac{\ELL}{2} \) \\
\sqrt{q_h} \cos\( \sqrt{q_h} \frac{\ELL}{2} \) & \sqrt{q_l}\cos\( \sqrt{q_l} \frac{\ELL}{2} \) & - \sqrt{q_h}\sin\( \sqrt{q_h} \frac{\ELL}{2} \) & - \sqrt{q_l}\sin\( \sqrt{q_l} \frac{\ELL}{2} \) \\
\sqrt{q_h} \cos\( \sqrt{q_h} \frac{\ELL}{2} \) & \sqrt{q_l}\cos\( \sqrt{q_l} \frac{\ELL}{2} \) & \sqrt{q_h}\sin\( \sqrt{q_h} \frac{\ELL}{2} \) & \sqrt{q_l}\sin\( \sqrt{q_l} \frac{\ELL}{2} \)
\end{array} \).
\end{align}
\end{subequations}
Adding the first two and subtracting the second two of these gives
\begin{align*}  
 c_3\cos\( \sqrt{q_h} \frac{\ELL}{2} \) + c_4 \cos\( \sqrt{q_l} \frac{\ELL}{2} \)&=0,\label{eqn:2x2-a}
 &
 c_3\sqrt{q_h}\sin\( \sqrt{q_h} \frac{\ELL}{2} \)+ c_4 \sqrt{q_l}\sin\( \sqrt{q_l} \frac{\ELL}{2} \)&=0.
\end{align*}
From this we can conclude that $c_3=0=c_4$ provided that
\begin{equation}\label{eqn:cB}
c_B \( \ELL,\gamma \) \equiv \cos\( \sqrt{q_h} \frac{\ELL}{2} \) \sqrt{q_l}\sin\( \sqrt{q_l} \frac{\ELL}{2} \) - 
\cos\( \sqrt{q_l} \frac{\ELL}{2} \) \sqrt{q_h}\sin\( \sqrt{q_h} \frac{\ELL}{2} \) \neq 0.
\end{equation}
To conclude the argument, we first note that \eqref{eqn:icond} can be recovered from \eqref{eqn:ele} and \eqref{eqn:cc} and hence also from \eqref{eqn:ode}, \eqref{eqn:cc} and \eqref{eqn:add}. Using \eqref{eqn:dans} in \eqref{eqn:icond} gives
\begin{equation}
\dfrac{c_3}{\sqrt{q_h}}\sin\( \sqrt{q_h} \frac{\ELL}{2} \) + \dfrac{c_4}{\sqrt{q_l}}\sin\( \sqrt{q_l} \frac{\ELL}{2} \) = \overline{y},  \label{eqn:4x4-e}
\end{equation}
from which $\overline{y}=0$ follows immediately.
\end{appendix}

\bibliographystyle{siamplain}
\bibliography{pkm1_refs}

\end{document}